\documentclass[12pt]{article}
\usepackage{amsmath,amssymb,epsfig}
\makeatletter \@addtoreset{equation}{section} \makeatother
\renewcommand{\theequation}{\thesection.\arabic{equation}}
\newcommand{\ba}{\begin{array}}
\newcommand{\ea}{\end{array}}
\newcommand{\beq}{\begin{equation}}
\newcommand{\eeq}{\end{equation}}
\newcommand{\bea}{\begin{eqnarray}}
\newcommand{\eea}{\end{eqnarray}}

\def\bce{\begin{center}}
\def\ece{\end{center}}

\def\nonu{\nonumber}

\def\pa{\partial}

\def\be{\beta}

\def\de{\delta}

\def\ep{\epsilon}

\def\diag{\mathop{\rm diag}}

\def\eps6{{\displaystyle \mathop{\epsilon}^{6}}{}}

\def\nab6{{\displaystyle \mathop{\nabla}^{6}}{}}

\def\0{{\sst{(0)}}}
\def\1{{\sst{(1)}}}
\def\2{{\sst{(2)}}}
\def\3{{\sst{(3)}}}
\def\4{{\sst{(4)}}}
\def\5{{\sst{(5)}}}
\def\6{{\sst{(6)}}}
\def\7{{\sst{(7)}}}
\def\8{{\sst{(8)}}}

\def\ba{\begin{array}}
\def\ea{\end{array}}
\def\beq{\begin{equation}}
\def\eeq{\end{equation}}
\def\be{\begin{equation}}
\def\ee{\end{equation}}

\def\diag{\mathop{\rm diag}}

\def\eps{\epsilon}

\def\ba{\begin{array}}
\def\ea{\end{array}}
\def\beq{\begin{equation}}
\def\eeq{\end{equation}}
\def\be{\begin{equation}}
\def\ee{\end{equation}}

\def\diag{\mathop{\rm diag}}

\def\eps{\epsilon}

\def\eps6{{\displaystyle \mathop{\epsilon}^{6}}{}}

\def\nab6{{\displaystyle \mathop{\nabla}^{6}}{}}

\newcommand{\bean}{\begin{eqnarray*}}
\newcommand{\eean}{\end{eqnarray*}}

\begin{document}
\thispagestyle{empty} \addtocounter{page}{-1}
   \begin{flushright}
\end{flushright}

\vspace*{1.3cm}
  
\centerline{ \Large \bf  The Eleven-Dimensional Uplift of   }
\vspace{.3cm} 
\centerline{ \Large \bf Four-Dimensional  Supersymmetric RG Flow }
\vspace*{1.5cm}
\centerline{{\bf Changhyun Ahn }
} 
\vspace*{1.0cm} 
\centerline{\it  
Department of Physics, Kyungpook National University, Taegu
702-701, Korea} 
\vspace*{0.8cm} 
\centerline{\tt ahn@knu.ac.kr 
} 
\vskip2cm

\centerline{\bf Abstract}
\vspace*{0.5cm}

The squashed and stretched 7-dimensional internal metric 
preserving $U(1) \times U(1) \times U(1)_R$ symmetry possesses
an Einstein-Kahler 2-fold which is a
base manifold of 5-dimensional Sasaki-Einstein $L^{p, q, r}$ space.   
The $r$(transverse to the domain wall)-dependence 
of the two 4-dimensional  supergravity fields,  that play the role of 
geometric parameters for squashing and
stretching, makes 
the 11-dimensional 
Einstein-Maxwell equations consistent not only at the two critical
points but also along the whole ${\cal N}=2$ supersymmetric RG flow connecting them.  
The Ricci tensor of the solution 
has common feature with the previous three
11-dimensional solutions.  
The 4-forms preserve only $U(1)_R$ symmetry for other generic parameters of
the metric.
We find an exact solution to the 11-dimensional Einstein-Maxwell
equations corresponding to the lift of the 4-dimensional supersymmetric RG flow. 

\baselineskip=18pt
\newpage
\renewcommand{\theequation}
{\arabic{section}\mbox{.}\arabic{equation}}

\section{Introduction}

The low energy limit of $N$ M2-branes at 
${\bf C}^4/{\bf Z}_k$ singularity is described
in the
${\cal N}=6$ $U(N) \times U(N)$ 
Chern-Simons matter theory
with level $k$ in 3-dimensions \cite{ABJM}.
For $k=1, 2$, the enhanced ${\cal N}=8$ supersymmetry
is preserved. 
The matter contents and the superpotential  of this theory
are the same as the ones in the theory of 
D3-branes at the conifold \cite{KW}. 
The RG flow of the 3-dimensional 
theory can be obtained from the 4-dimensional gauged
supergravity  via AdS/CFT 
correspondence \cite{Maldacena}. 
The holographic supersymmetric
RG flow connecting the maximally supersymmetric point 
to ${\cal N}=2$ $SU(3) \times U(1)_R$ point has been studied in 
\cite{AP,AW} while
those from this maximally supersymmetric point 
to
${\cal N}=1$ $G_2$ point has been studied in 
\cite{AW,AI}.
The 11-dimensional M-theory uplifts of these
have been found in \cite{CPW,AI}  by solving the
Einstein-Maxwell equations in 11-dimensions sometime ago.

The mass deformed $U(2) \times U(2)$
Chern-Simons matter theory with level $k=1, 2$ 
preserving the above global $SU(3) \times U(1)_R$ symmetry has been studied 
in \cite{Ahn0806n2,BKKS} while
the mass deformation for this theory preserving $G_2$
symmetry  has been described and 
the further nonsupersymmetric 
RG flow equations preserving two $SO(7)$ symmetries 
have been discussed in \cite{Ahn0806n1}.  
The holographic
RG flow equations connecting  $G_2$ point 
to  $SU(3) \times U(1)_R$ point have been 
found in \cite{BHPW}.  
Moreover, the other holographic supersymmetric
RG flows have been 
studied and  
further developments on  
the 4-dimensional gauged supergravity have been done
in \cite{AW09}. 
The spin-2 Kaluza-Klein modes around 
a warped product of $AdS_4$ and a seven-ellipsoid which has 
above global $G_2$ symmetry are discussed in \cite{AW0907}. 
The gauge dual with the symmetry of
$SU(2) \times SU(2) \times U(1)_R$ for the 11-dimensional lift of 
$SU(3) \times U(1)_R$-invariant solution 
in 4-dimensional supergravity 
is described in \cite{AW0908}.
Recently, the 11-dimensional description preserving $SU(2) \times U(1)
\times U(1)_R$ symmetry is found in \cite{Ahn0909}.
They have common $U(1)_R$ factor.   

When
the 11-dimensional theory from the 4-dimensional gauged
supergravity is constructed, the various
11-dimensional solutions will occur even though the flow equations
characterized by the 4-dimensional supergravity fields 
 are
the same. Since the 4-dimensional flow
equations are related to the ${\cal N}=2$ supersymmetry through $U(1)_R$
symmetry, other types of 
11-dimensional solutions with common 4-dimensional flow equations will
be possible. 
The invariance of 11-dimensional metric and 4-forms determines 
each global symmetry. Sometimes the 4-forms restrict to the global
symmetry the metric contains and break into the smaller symmetry group. 
In \cite{CPW}, the two different 11-dimensional solutions
where the first has ${\bf CP}^2$ space with $SU(3) \times U(1)_R$
symmetry
and the second has ${\bf CP}^1
\times {\bf CP}^1$ space with 
$SU(2) \times SU(2) \times U(1)_R$ symmetry are found. 
Furthermore, the third 11-dimensional solution \cite{Ahn0909}
with a single ${\bf CP}^1$ space with $SU(2) \times U(1) \times U(1)_R$ 
symmetry is described.
The Ricci tensor for these three
solutions with orthonormal frame basis has same value 
by assuming that the 4-dimensional supergravity fields satisfy the same
equations of motion discovered by \cite{AP} sometime ago. 
That is, the same flow equations in 4-dimensions provide
three different 11-dimensional solutions to the equations of the
motion in 11-dimensional supergravity: $SU(3) \times U(1)_R$ flow
corresponding to homogeneous five-sphere,
$SU(2) \times SU(2) \times U(1)_R$ flow corresponding to homogeneous
 $T^{1,1}$
space 
and $SU(2) \times U(1) \times
U(1)_R$ flow corresponding to cohomogeneity one $Y^{p,q}$ space.

In this paper, we construct 
a new 11-dimensional solution preserving the $U(1)_R$ symmetry. 
By assuming that the 4-dimensional $AdS_4$ supergravity
fields satisfy the supersymmetric RG flow equations,  
we find out the correct deformed 7-dimensional internal space possessing
the correct global symmetry. 
By realizing that the 5-dimensional Sasaki-Einstein $Y^{p,q}$ space
can be generalized to the 5-dimensional Sasaki-Einstein $L^{p, q, r}$ 
space \cite{CLPP,MS},
we focus on this cohomogeneity two $L^{p, q, r}$ space. 
When the parameters of the metric satisfy 
$\alpha=\beta$, the $L^{p, q, r}$ space is nothing but $Y^{p, q}$ 
space and moreover the isometry of $L^{p, q, r}$ is given by the 
$U(1) \times U(1) \times U(1)_R$ symmetry. The flow diagrams for 
four different five-dimensional Sasaki-Einstein spaces $({\bf S}^5, 
T^{1,1}, Y^{p,q}, L^{p,q,r})$ are given in  \cite{AV}.

We will start with the round
compactification in terms of $U(1)$-fibration over the Einstein-Kahler
3-fold,
squash this Einstein-Kahler base ellipsoidally, stretch the
$U(1)$ fiber, and introduce 3-form tensor gauge potential
proportional to the volume form on the base, in the spirit of \cite{CPW}. 
Basically the structure of 3-form from the
triple wedge product between the orthonormal frames looks similar to
the previous three cases. The overall functional dependence on the $AdS_4$
supergravity fields and the exponential factors corresponding to
the unbroken $U(1)$ symmetries can be determined by solving the 11-dimensional
Einstein-Maxwell equations directly.

In section 2, starting with the two parts of $L^{p, q, r}$ space metric,
$U(1)$ bundle and the Einstein-Kahler 2-fold, 
we embed them inside of the squashed and stretched  
7-dimensional internal space appropriately.  Then we determine the
full 11-dimensional metric with the correct warp factor. 
Assuming that the two supergravity fields satisfy the domain
wall solutions, we compute the Ricci tensor in this background completely.  
Surprisingly, the Ricci tensor with orthonormal frame basis has the 
same value in previous three cases found before.
For the 4-form field strengths, we make an ansatz by writing the
three pieces: the overall function, the exponential function with
$U(1)$'s and the triple wedge product between the orthonormal frames.
Finally, we determine the solution for 
the 11-dimensional Einstein-Maxwell equations.

In section 3, we summarize the results of this paper and present some
future directions.

In the Appendix, we present the detailed expressions for the 
4-form field strengths.

\section{An ${\cal N}=2$ supersymmetric flow in an 11-dimensional theory}

Let us describe the 11-dimensional metric. 
The 3-dimensional metric is given by 
$\eta_{\mu\nu}=(-,+,+)$, the radial variable $r$ that is the fourth
coordinate 
is  transverse to
the domain wall, and the scale factor $A(r)$ in 3-dimensional metric 
behaves linearly in $r$ at UV
and IR regions.
The 4-dimensional spacetime metric in the 11-dimensions contains
a warp factor  $\Delta(r,\mu)$ that depends on the $\mu$ which is 
the fifth coordinate as well as the $r$. The 7-dimensional 
internal metric depends on the 4-dimensional 
supergravity fields $(\rho, \chi)$ and the warp factor in the full 11-dimensions.   
Explicitly, we have the
11-dimensional metric as follows \cite{CPW,AI02-1}:
\bea
ds_{11}^2 =\Delta(r,\mu)^{-1} \, \left(dr^2 +e^{2 A(r)}
\, \eta_{\mu\nu}\, dx^\mu dx^\nu \right)+ L^2 \,
\Delta(r,\mu)^{\frac{1}{2}} \, ds_7^2(\rho,\chi),
\label{11dmetric}
\eea
where $L$ is a radius of round seven-sphere.

In 4-dimensions, there are two critical points, 
${\cal N}=8$ $SO(8)$ critical
point and ${\cal N}=2$
$SU(3) \times U(1)_R$ critical point. 
At these points, the derivatives of superpotential $W(\rho, \chi)$
with respect to the supergravity fields vanish.
Let us recall that the supergravity fields $(\rho, \chi)$ and the
scale function $A$ appearing in the metric (\ref{11dmetric})
satisfy the supersymmetric RG flow equations \cite{AP}:
\bea
\frac{d \rho}{d r} & = & \frac{1}{8L \, \rho} \, \left[
  (\cosh(2\chi) +1) + \rho^8\, (\cosh(2\chi)-3) \right],
\nonu \\
\frac{d \chi}{d r} & = & \frac{1}{2L \, \rho^2} \,
(\rho^8-3) \,
\sinh(2\chi),
\nonu \\
\frac{d A}{d r} & = & \frac{1}{4L \, \rho^2} \, 
\left[ 3
(\cosh(2\chi)+1) -\rho^8 (\cosh(2\chi)-3)\right].
\label{domain}
\eea

As explained in the introduction,
let us consider the 4-dimensional 
Einstein-Kahler 2-fold which lives in the five-dimensional $L^{p, q, r}$
space \cite{CLPP,MS}. 
The one-form containing 
the $U(1)$ bundle over this Einstein-Kahler 2-fold consists of 
two parts as follows \cite{CPW,AI02-1}:
\bea
\omega = \frac{1}{2} \, \sin (2\mu) \left[-\rho(r)^{-4} \, d \gamma +
  \rho(r)^4 \, (u, J d u)\right],
\label{omega}
\eea
where the 8-dimensional vector 
$u=(u^1, \cdots, u^6, 0, 0)$ parametrizes a unit
five-sphere, $\gamma$ is an 11-th coordinate 
and $J$ is the Kahler form that has
$J_{12}=J_{34}=J_{56}=J_{78}=1$ explicitly.

One needs to know $(u, J d u)$ in (\ref{omega}) corresponding to the $U(1)$
bundle over the Einstein-Kahler 2-fold.
Let us recall the metric for 
the 5-dimensional Sasaki-Einstein space $L^{p, q, r}$ used in 
\cite{MS}
\bea
&& ds^2_{L^{p,q,r}}   =  
ds_{EK(2)}^2 +  \left[ d \tau -2(\xi+\eta) \, d \phi -
2\xi \eta \, d \psi \right]^2,
\label{lpqr}
\eea
with the  Einstein-Kahler 2-fold \footnote{One can also consider the
  metric by \cite{CLPP} but the trigonometric functions on the 
angle $\theta$ appear in the metric and this makes the Ricci tensor be
complicated expressions. However, the parametrization of \cite{MS} we
use in this paper can make the metric (\ref{lpqr}) 
take the form of \cite{CLPP}. So we take the convention of \cite{MS}.}
\bea
ds_{EK(2)}^2
&=& 
 \frac{(\eta-\xi)}{2 G(\eta)} \, d \eta^2 +
 \frac{(\eta-\xi)}{2 F(\xi)} \, d \xi^2   
+ \frac{2F(\xi)}{(\eta-\xi)} \left(d \phi + 
\eta\, d \psi \right)^2
+\frac{2G(\eta)}{(\eta-\xi)} \left(d \phi + 
\xi \, d \psi \right)^2
\label{EK}
\eea
where 
the $\xi$ and $\eta$-dependent cubic functions with parameters
$\alpha,\beta,\mu_1$ 
are given by
\bea
F(\xi) \equiv 2\xi(\alpha-\xi)(\alpha-\beta-\xi), \qquad 
G(\eta)
\equiv -2\eta(\alpha-\eta)(\alpha-\beta-\eta)-2 \mu_1.
\label{FG}
\eea
It is obvious that the form in the last term of (\ref{lpqr}) provides the
Kahler 2-form and satisfies
\bea
d  \left[  -2(\xi+\eta) \, d \phi -
2\xi \eta \, d \psi \right] = -2\left[ d \xi \wedge (d \phi + \eta \, d
\psi)+
d \eta \wedge (d \phi + \xi \, d \psi) \right]. 
\label{Kahler2}
\eea

Therefore,  
one identifies  $(u, J d u)$ with the $U(1)$ bundle over 
the Einstein-Kahler 2-fold as follows:
\bea
(u, J d u) = d \tau -2(\xi+\eta) \, d \phi -
2\xi \eta \, d \psi.
\label{uJdu}
\eea 
What about the $U(1)$ Hopf fiber $(x, J d x)$
on ${\bf CP}^3$  where
$x=(x^1,\cdots,x^8)$ is a vector on 8-dimensional space
in terms of
$(u, J d u)$?
This $U(1)$ Hopf fiber becomes \cite{CPW,AI02-1}
\bea
(x, J d x) = \cos^2\mu \, (u, J d u) + \sin^2\mu \, d \gamma.
\label{xJdx}
\eea

The 7-dimensional internal space metric $ds_7^2(\rho, \chi)$ appearing
in the metric (\ref{11dmetric}) can be 
written as \footnote{The parameter $\xi$ in \cite{CPW} is replaced by a
capital letter $\Xi$ in order not to confuse with the coordinate $\xi$ in (\ref{EK}). }
\bea
&& ds_7^2   =  \rho(r)^{-4}  \Xi^2  d\mu^2 + \rho(r)^2  \cos^2\mu
ds^2_{EK(2)} + \Xi^{-2} \omega^2 +\Xi^{-2} \cosh^2\chi(r)  (x, J
d x)^2.
\label{7d}
\eea
By substituting  the metric (\ref{EK}) for the Einstein-Kahler 2-fold
into the second term of (\ref{7d}),
plugging the 1-form (\ref{omega}) together with (\ref{uJdu}) into the third
term of (\ref{7d}) and substituting the
$U(1)$ Hopf fiber (\ref{xJdx}) into the last term of (\ref{7d}), 
finally one obtains the final 7-dimensional
internal metric preserving $U(1) \times U(1) \times U(1)_R$ symmetry
as follows:
\bea
&& ds_7^2(\rho, \chi) =    \rho(r)^{-4} \, \Xi(r,\mu)^2 \, d\mu^2 
\label{7dmetric}
 \\
&& +  
\rho(r)^2 \, \cos^2\mu
\, \left[  
 \frac{(\eta-\xi)}{2 G(\eta)} \, d \eta^2 +
 \frac{(\eta-\xi)}{2 F(\xi)} \, d \xi^2   
+ \frac{2F(\xi)}{(\eta-\xi)} \left(d \phi + 
\eta\, d \psi \right)^2
+\frac{2G(\eta)}{(\eta-\xi)} \left(d \phi + 
\xi \, d \psi \right)^2
\right] 
\nonu \\
&&+  \Xi(r,\mu)^{-2} \,  \frac{1}{4} \, \sin^2 (2\mu) \,
\left(-\rho(r)^{-4} \, d \gamma +
  \rho(r)^4 \,   \left[  d \tau -2(\xi+\eta) \, d \phi -
2\xi \eta \, d \psi \right]
\right)^2
\nonu \\
&& +  \Xi(r,\mu)^{-2} \, \cosh^2\chi(r) \, \left( \sin^2\mu \, d \gamma +
\cos^2\mu \,   \left[ d \tau -2(\xi+\eta) \, d \phi -
2\xi \eta \, d \psi  \right] \right)^2.
\nonu
\eea
Here one has 
\bea
\Xi(r,\mu) = \frac{\sqrt{X(r,\mu)}}{\rho(r)}, \qquad
X(r,\mu) \equiv \cos^2\mu + \rho(r)^8 \, \sin^2\mu.
\label{X}
\eea 
The nontrivial squashing characterized by $\rho(r)$ deforms the metric 
on the ${\bf CP}^3$ 
and moreover rescales the Hopf fiber which appears in the last line of
(\ref{7dmetric}). The stretching is characterized by $\chi(r)$.
There exists $U(1) \times U(1)$ symmetry
from the structure of Einstein-Kahler 2-fold in $ds^2_{EK(2)}$.
These two $U(1)$ symmetries are generated by the 8-th coordinate 
$\phi$ and 9-th coordinate $\psi$.
The combined two $U(1)$ symmetries by the 10-th coordinate $\tau$ and 
11-th coordinate $\gamma$
will provide a single $U(1)_R$ symmetry relevant to the
${\cal N}=2$ supersymmetry later. 

For $\mu=0$, the 7-dimensional metric (\ref{7dmetric}) 
reduces to the following metric on moduli space for the M2-brane probe
\bea
\rho(r)^2 ds^2_{L^{p,q, r}} +\rho(r)^2 \sinh^2 \chi(r) \,
 \left[ d \tau -2(\xi+\eta) \, d \phi -
2\xi \eta \, d \psi  \right]^2,
\label{modu}
\eea
where the metric for ${L^{p, q, r}}$ is given by (\ref{lpqr}).  
The function  $\sinh^2 \chi(r)$ in (\ref{modu}) plays the role of a
stretching of the $U(1)$-fiber. Then for this particular
coordinate $\mu=0$ there exists 
a stretched $L^{p, q, r}$ space.

One obtains the following set of orthonormal frames for the
11-dimensional metric (\ref{11dmetric}) as follows:
\bea
e^1  & = & -\Delta(r,\mu)^{-\frac{1}{2}} \, e^{A(r)} \, d x^1,
\qquad
e^2   =  \Delta(r,\mu)^{-\frac{1}{2}} \, e^{A(r)} \, d x^2,
\qquad
e^3   =  \Delta(r,\mu)^{-\frac{1}{2}} \, e^{A(r)} \, d x^3,
\nonu \\
e^4   &  = &  \Delta(r,\mu)^{-\frac{1}{2}}  \, d r,
\qquad
e^5    =   L \,
\Delta(r,\mu)^{\frac{1}{4}} \, \frac{\sqrt{X(r,\mu)}}{\rho(r)^3} \, d \mu,
\nonu \\
 e^6  &  = &  L\,
\Delta(r,\mu)^{\frac{1}{4}} \,\rho(r) \cos \mu \, 
\sqrt{\frac{\eta-\xi}{2G(\eta)}} \, d \eta,
\nonu \\
 e^7 & = & L\,
\Delta(r,\mu)^{\frac{1}{4}} \,\rho(r) \cos \mu \,
\sqrt{\frac{\eta-\xi}{2F(\xi)}} \, d \xi,
\nonu \\
 e^8  & = &  L\,
\Delta(r,\mu)^{\frac{1}{4}} \,\rho(r) \cos \mu \, 
\sqrt{\frac{2F(\xi)}{\eta-\xi}} \, \left[ d \phi + 
\eta \, d \psi \right],
\nonu \\
 e^9 & = & L\,
\Delta(r,\mu)^{\frac{1}{4}} \,\rho(r) \cos \mu \, 
\sqrt{\frac{2G(\eta)}{\eta-\xi}} \, \left[ d \phi + 
\xi \, d \psi \right],
\label{11frames}
\\
 e^{10} & = & L\,
\Delta(r,\mu)^{\frac{1}{4}} \, \frac{\rho(r)}{\sqrt{X(r,\mu)}} \, \frac{1}{2} \, \sin(2\mu) \, 
 \left( - \rho(r)^{-4} \, d \gamma
  + \, \rho(r)^4\,   \left[  d \tau -2(\xi+\eta) \, d \phi -
2\xi \eta \, d \psi  \right] \right),
\nonu
\\
 e^{11} & = & L\,
\Delta(r,\mu)^{\frac{1}{4}} \,  \frac{\rho(r) \, \cosh \chi(r)}{\sqrt{X(r,\mu)}} \, 
 \left( \sin^2\mu \, d \gamma + 
 \cos^2 \mu \,   \left[   d \tau -2(\xi+\eta) \, d \phi -
2\xi \eta \, d \psi \right] \right),
\nonu
\eea
where the warp factor is 
\bea
\Delta(r,\mu) = \frac{\rho(r)^{\frac{4}{3}}}{ X(r,\mu)^{\frac{2}{3}}
  \, 
\cosh^{\frac{4}{3}} \chi(r) }.
\label{Delta}
\eea

The
Einstein-Maxwell equations are given by \cite{CJS,DNP}
\bea
R_{M}^{\;\;\;N} & = & \frac{1}{3} \,F_{MPQR} F^{NPQR}
-\frac{1}{36} \de^{N}_{M} \,F_{PQRS} F^{PQRS},
\nonu \\
\nabla_M F^{MNPQ} & = & -\frac{1}{576} \,E \,\ep^{NPQRSTUVWXY}
F_{RSTU} F_{VWXY},
\label{fieldequations}
\eea
where the covariant derivative $\nabla_M$ 
on $F^{MNPQ}$
is given by 
$E^{-1} \pa_M ( E F^{MNPQ} )$ together with elfbein determinant 
$E \equiv \sqrt{-g_{11}}$. The epsilon tensor 
 $\ep_{NPQRSTUVWXY}$ with lower indices is purely numerical.
All the indices in (\ref{fieldequations}) are based on the coordinate basis.
For given 11-dimensional metric (\ref{11dmetric}) together with
(\ref{Delta}) and (\ref{7dmetric})
or (\ref{11frames}), 
the next step is to find the solution for (\ref{fieldequations}).

Let us describe the 11-dimensional solution at two critical points and
after that along the whole RG flow connecting them.

$\bullet$ At  the UV fixed point 

At this critical point
\bea
\rho(r) = 1, \qquad \chi(r) =0,
\label{rhochi}
\eea
one recovers the maximally supersymmetric ${\cal N}=8$ 
$AdS_4 \times {\bf S}^7$ solution \cite{DNP} and 
the Ricci tensor has the form
\bea
R_M^{\,\,N} =\frac{6}{L^2} \diag (-2, -2, -2, -2,
1, 1, 1, 1, 1, 1, 1 ).
\nonu
\eea 
The 3-form gauge field with 
3-dimensional M2-brane indices is defined by  \cite{CPW}
\bea
A^{(3)} = \frac{1}{2} e^{\frac{6r}{L}} \, d x^1
\wedge d x^2 \wedge d x^3.
\label{A3}
\eea
At the UV end, the scale function $A(r)$ behaves as
$\frac{2}{L} r$ and one obtains the only nonzero component for the 4-form 
as $ F_{1234} =-\frac{18}{L}$ \cite{FR}.

$\bullet$ At  the IR fixed point 

At this critical point
\bea
\rho(r) = 3^{\frac{1}{8}}, \qquad \chi(r) =\frac{1}{2} \cosh^{-1} 2,
\label{rhochi1}
\eea
the function $A(r)$ behaves as
$\frac{3^{\frac{3}{4}}}{L} r$($\hat{L} \equiv
3^{-\frac{3}{4}} L$), then one writes down the 3-form gauge field 
as follows \cite{CPW}:
\bea
A^{(3)} = \frac{3^{\frac{3}{4}}}{4} e^{\frac{3r}{\hat{L}}} \, d x^1
\wedge d x^2 \wedge d x^3 + C^{(3)} + (C^{(3)})^{\ast}.
\label{a3}
\eea
Since the Kahler form in (\ref{Kahler2}) contains 
$e^6 \wedge e^9$ and $e^7 \wedge e^8$, this leads to
the natural basis of the one-forms and the ${\bf CP}^3$ factor for
$\rho=1$ and $\chi=0$ (\ref{rhochi}) has also $e^5$ and $e^{10}$ which can be
combined together.
In fact, we find 
\bea
C^{(3)} = -\frac{1}{4} \sinh \chi(r) \, e^{-i[4(\beta-2\alpha)\phi+
2\alpha(\beta-\alpha)\psi+3\tau +\gamma]} \, (e^5+ i e^{10}) 
\wedge (e^6 + i e^9) \wedge (e^7 + i e^8).
\label{c3}
\eea
Although the structure of triple wedge product in (\ref{c3}) 
between the orthonormal basis looks very similar to the previous 
constructions with $SU(3) \times U(1)_R$ 
symmetry \cite{CPW}, $SU(2) \times SU(2) \times
U(1)_R$ symmetry \cite{CPW} or 
$SU(2) \times U(1) \times U(1)_R$ symmetry \cite{Ahn0909}, 
the functional behavior of the
exponential function in 3-form behave differently.   

The Ricci tensor has only two nonvanishing off-diagonal
components:$R_{10}^{\,11}$ and $R_{11}^{\,10}$. 
It turns out the Ricci tensor is identical to the one with $SU(3)
\times U(1)_R$ symmetry \cite{CPW}, 
the one with $SU(2) \times SU(2) \times U(1)_R$ symmetry \cite{CPW,AW0908} or
the one with $SU(2) \times U(1) \times U(1)_R$ symmetry \cite{Ahn0909}.
That is, the Ricci tensor for four cases 
has same value(in the frame basis) at the IR critical point (\ref{rhochi1}).
They are given by \cite{AW0908,Ahn0909}
\bea
R_1^{\, 1} & = & -\frac{(55-32 \cos 2\mu + 3 \cos 4\mu) }
{3 \cdot 2^{\frac{1}{3}} \,\sqrt{3} \, \hat{L}^2\,
  (2-\cos 2\mu )^{\frac{8}{3}}} =R_2^{\, 2} =R_3^{\, 3}=R_4^{\, 4}
= -2 R_6^{\, 6} = -2 R_7^{\, 7} = -2
R_8^{\, 8}=-2 R_9^{\, 9},
\nonu \\
R_5^{\, 5} & = & \frac{(29-16 \cos 2\mu) }
{3 \cdot 2^{\frac{1}{3}} \,\sqrt{3} \,\hat{L}^2\,
  (2-\cos 2\mu )^{\frac{8}{3}}} = R_{10}^{\, 10}, \qquad
R_{10}^{\, 11} =  -\frac{2 \cdot 2^{\frac{1}{6}} \sin 2\mu }
{\sqrt{3} \,\hat{L}^2\,
  (2-\cos 2\mu )^{\frac{5}{3}}} = R_{11}^{\,\,10},
\nonu \\
R_{11}^{\, 11} & = & \frac{(80-64 \cos 2\mu +9 \cos 4 \mu) }
{3 \cdot 2^{\frac{1}{3}} \,\sqrt{3} \,\hat{L}^2\,
  (2-\cos 2\mu )^{\frac{8}{3}}}.
\label{Ricci}
\eea
They depend on only the fifth coordinate $\mu$. 
One transforms the Einstein equation with
coordinate basis into the one with frame basis
via (\ref{11frames}). 
By comparing the $(10, 11)$  component of Einstein
equation,
the coefficients for the angles $\tau$ and $\gamma$ which are equal to $-3$ 
and $-1$ and the overall coefficient of
3-form that is $-\frac{1}{4}$ are completely fixed. 
Moreover, the $(10, 9)$ component of 
right hand side of Einstein equation is nonzero but 
the corresponding $R_{10}^{\,9}$ from (\ref{Ricci}) vanishes. This implies 
that the coefficient of $\phi$ should be $8\alpha-4\beta$ and the coefficient of
$\psi$ should be $2\alpha^2-2\alpha \beta$ in the exponent of 3-form (\ref{c3}).
Then there exists a $U(1)$ symmetry generated by the angle $\phi$ or
$\psi$
with particular conditions on $\alpha$ and $\beta$.
Either $(2\alpha-\beta)=0$ or $\alpha(\alpha- \beta)=0$.

The internal part of $F^{(4)}$ can be written as
$ d  C^{(3)} + d (C^{(3)})^{\ast}$ with (\ref{c3}) and
the antisymmetric 
tensor fields can be obtained  from 
$F^{(4)} = d A^{(3)}$ with (\ref{a3}).
It turns out that the antisymmetric 
field strengths have the following nonzero components
in the orthonormal frame basis: 
\bea
F_{1234}  & = & -\frac{3 \cdot  2^{\frac{1}{3}} \cdot 3^{\frac{3}{4}}}
{\hat{L}(2-\cos2\mu)^{\frac{4}{3}}}, \nonu \\
F_{579\,10} + i \, F_{567\,10}  & = &  \frac{2^{\frac{1}{3}} \cdot 3^{\frac{3}{4}}
   \sin2\mu}
{\hat{L}(2-\cos2\mu)^{\frac{4}{3}}} \, e^{i \delta} =-F_{568\,10} + i
\, F_{589\,10},
\nonu \\
F_{579\,11} + i \, F_{567\,11} & = & -\frac{2^{\frac{5}{6}} \cdot 3^{\frac{3}{4}} }
{\hat{L}(2-\cos
  2\mu)^{\frac{1}{3}}} \, e^{i \delta} =-F_{568\,11} + i \, F_{589\,11},
\nonu \\
& = &
-F_{67\,10\,11} + i \, F_{79\,10\,11} = -F_{89\,10\,11} - i \, F_{68\,10\,11},
\label{F4}
\eea
where we introduce the exponent appearing in (\ref{c3}) as a single variable
\bea
\delta \equiv 4(\beta-2\alpha)\phi+
2\alpha(\beta-\alpha)\psi+3\tau +\gamma.
\label{delta} 
\eea
The angle-dependences for $\phi, \psi, \tau$ and $\gamma$ appear via (\ref{delta}). 
One can make the four $U(1)$ symmetries 
generated by these angles 
which preserve the $\delta$.
These 4-forms break the $U(1) \times U(1) \times U(1)
\times U(1)$ into $U(1)_R$. 
After substituting (\ref{F4}) into 
the right hand side of Einstein equation (\ref{fieldequations})
with frame basis (\ref{11frames}) 
one reproduces the one for $SU(3) \times U(1)_R$ symmetry case
\cite{CPW},  $SU(2) \times SU(2) \times U(1)_R$, or $SU(2) \times U(1)
\times U(1)_R$  \cite{Ahn0909} exactly. 
This feature is also expected because the Ricci
tensor for four independent cases is identical to each other.
That is, the 4-forms themselves are different from each other but 
their quadratic combinations 
appearing in the right hand side of Einstein equation
are the same.
Note that the 4-form given in (\ref{F4}) looks very similar to 
the one of $SU(2) \times SU(2) \times U(1)_R$ symmetry case 
\cite{CPW,AW0908} or $SU(2)
\times U(1) \times U(1)_R$ symmetry case \cite{Ahn0909}: same
independent components(up to signs). 

$\bullet$ Along the  the RG flow

Now let us consider the whole RG flow.
For solutions with varying scalars, the ansatz for the 4-form field
strength will be more complicated. 
We apply the correct ansatz for the 11-dimensional 3-form gauge
field by acquiring the $r$-dependence of the 4-dimensional 
supergravity scalars and
derive the 11-dimensional Einstein-Maxwell equations
corresponding to the $U(1)_R$-invariant RG flow.

Let us take the 3-form ansatz as follows: 
\bea
A^{(3)} = \widetilde{W}(r,\mu) \, e^{3A(r)} \, d x^1
\wedge d x^2 \wedge d x^3 + C^{(3)} + (C^{(3)})^{\ast},
\label{a3flow}
\eea
where $ C^{(3)}$ is given by (\ref{c3}) as before. 
One puts an arbitrary function $f(\rho,\chi)$ in front of this 3-form
at the beginning.
One obtains the Ricci tensor from the 11-dimensional 
metric (\ref{11dmetric})
when the supergravity fields $(\rho, \chi)$ vary with respect to the
$r$-coordinate. They are exactly the same as the one in the Appendix A
of \cite{Ahn0909}.
The $(10, 11)$ component of Einstein equation determines the function
$f(\rho, \chi)$. One obtains $v f(v) +(1-v^2)f'(v)=0$
where $v \equiv \cosh\chi$. This implies that the solution
$f(v)$ is exactly the same as $\sinh\chi$ which appears in (\ref{c3}).

One determines the exact form for the geometric superpotential
introduced in (\ref{a3flow}).
Let us consider $(4, 4), (4, 5)$- and $(5, 5)$-components of the right
hand side of Einstein equation.
By eliminating $(\pa_{r}  \widetilde{W})^2 $  
from $(4,4)$- and $(5,5)$-components, one obtains
$\pa_{\mu} \widetilde{W}(r,\mu)$.
By integrating this with respect to the $\mu$ coordinate, one gets
$
\widetilde{W}(r,\mu)$ with undetermined function for the $r$. 
By making the correct ansatz for this function, one can determine it completely
from the $(4,5)$ component of Einstein equation. 
Actually,  the $(4, 4), (4, 5)$- and $(5, 5)$-components of the right
hand side of Einstein equation are exactly the same as 
the ones in \cite{Ahn0909}. 
Therefore, 
one obtains the final form for the geometric superpotential
as follows:
\bea
 \widetilde{W}(r,\mu) = \frac{1}{4\rho(r)^2} \left[ 
\left(\cosh2\chi(r) +1 \right) \cos^2 \mu - \rho(r)^8 \left( \cosh2\chi(r)-
 3 \right) \sin^2 \mu \right],
\label{geometric}
\eea
which is exactly the same as the one \cite{CPW} found in other three cases.
Note that this reproduces the UV value in (\ref{A3}) or the IR value
in (\ref{a3}). 

Comparing with the previous 4-form fields at the IR fixed point, 
the mixed 4-form fields $F_{\mu\nu\rho 5}, F_{4mnp}$ and $F_{45mn}$
where $\mu, \nu, \rho =1, 2, 3$ and $m,n, p =6, 7, \cdots, 11$ are
new if we look at the (\ref{F41}). We also present the 4-forms in the
coordinate basis in Appendix B.
For the checking of the remaining Maxwell equation
(\ref{fieldequations}), 
one needs to know
the elfbein determinant $E=\sqrt{-g_{11}}$ and it turns out that it is given by
\bea
E = 243 \cdot \, 3^{\frac{1}{4}} \, e^{3A(r)}
\, \hat{L}^7 \,  \rho(r)^{-\frac{4}{3}} \,
\cosh^{\frac{4}{3}} \chi(r)  \, (\eta-\xi) \, \cos^5\mu \,
 \, \sin \mu \,
\left( \cos^2\mu + \rho(r)^8 \, \sin^2\mu\right)^{\frac{2}{3}}.
\nonu
\eea
Moreover, the determinant of inverse metric is 
$g_{11}^{-1} = \epsilon^{123456789\,10\,11}=-E^{-2}$.
We have checked that all of the Maxwell equations of 
motion are indeed satisfied.

Thus we have constructed that the solutions (\ref{a3flow}), (\ref{c3}), and 
(\ref{geometric}) consist of an exact solution to the
11-dimensional supergravity characterized by bosonic field equations
(\ref{fieldequations}), 
provided that the deformation parameters
$(\rho, \chi)$ of the 7-dimensional internal space and the
domain wall amplitude $A$ develop in the $AdS_4$ radial direction   
along the RG flow (\ref{domain}).

For $\alpha=\beta$(after we go to the metric by \cite{CLPP}), then the metric of 
(\ref{lpqr}) leads to the standard metric of $Y^{p,q}=L^{p-q,p+q,p}$ space. 
For $p=q=r=1$, 
the metric provides the homogeneous $T^{1,1}$ space.
For $\mu_1=0$, the metric becomes
the round five-sphere metric \cite{CLPP}.
In the 11-dimensional view point, the four independent RG flows
characterized by 
\bea
{\bf S}^5-\mbox{flow} & : & SU(3) \times U(1)_R,  \nonu \\
T^{1,1}-\mbox{flow} & : & SU(2) \times SU(2) \times U(1)_R, \nonu \\
Y^{p,q}-\mbox{flow}& : & SU(2) \times U(1) \times U(1)_R, 
\nonu \\
L^{p, q, r}-\mbox{flow}& : & U(1) \times U(1)_R, \,\,\, U(1)_R, 
\label{four}
\eea
arrive at the IR
fixed point at which they have common Ricci tensor given in the
Appendix A of \cite{Ahn0909}.
Depending on their global symmetry, the internal 3-forms, in each case,  
have the right structures in the exponential function with common 
$\sinh\chi$-dependence. However, the 3-form in the M2-brane 
world-volume directions with the same geometric superpotential 
(\ref{geometric}) 
is common to four different solutions (\ref{four}). 
Although the 4-forms are different from 
each other completely, the squares of these 4-forms appearing in the
right hand  side of Einstein equation (\ref{fieldequations}) give
rise to the same expressions.  

\section{
Conclusions and outlook }

The solutions, characterized by (\ref{c3}), 
(\ref{a3flow}) and (\ref{geometric}) together with (\ref{11frames}), 
for 11-dimensional Einstein-Maxwell equations
corresponding to the ${\cal N}=2$ $U(1)_R$-invariant
RG flow in the 4-dimensional gauged supergravity are found. 
More explicitly, the Ricci tensor is given by the Appendix A of \cite{Ahn0909}
and the 4-forms are given by the Appendix B of this paper. 
These two quantities are the basic objects in the 11-dimensional
Einstein-Maxwell equations (\ref{fieldequations}). The 4-forms with
upper indices can be obtained from those with lower indices given in
the Appendix B via the 11-dimensional metric (\ref{11frames}).
Note that 
the $U(1) \times U(1) \times U(1)_R$ symmetry of 11-dimensional metric
breaks into $U(1)_R$ in the presence of 4-form field strengths for
general parameters $\alpha$ and $\beta$. One
can interpret the $AdS_4$
supergravity fields as the geometric parameters for
the 7-dimensional internal space and as long as the $r$-dependence of
these fields is controlled by the supersymmetric RG flow equations, 
the exact solution for the 11-dimensional field equations is
determined. 
Therefore
the $U(1)_R$-invariant
holographic RG flow is lifted to an ${\cal N}=2$ M2-brane flow in
M-theory.

It is an open  problem to find out what is corresponding dual gauge theory for the
11-dimensional background we have described in the context of AdS/CFT.
In \cite{GMSW1}, the higher dimensional analog of the 
5-dimensional $Y^{p,q}$ space was found. 
Then the partial resolution of the 7-dimensional space might be a
candidate for the dual gauge theory, along the line of \cite{FKR}. 
It would be interesting to study the other possibility 
where there exists a bigger $SU(3) \times U(1) \times U(1)_R$ symmetry
for the 11-dimensional lift of the same 4-dimensional RG flow
equations. 
The 11-dimensional lift of ${\cal N}=1$ $G_2$ invariant theory
was found in \cite{AI}. It is an open problem 
whether one can embed the appropriate Einstein-Kahler 2-fold inside of 
six-sphere  or whether one can replace other Einstein-Kahler 3-fold. 
As mentioned in the introduction, due to the flow from $G_2$ point to
$SU(3)\times U(1)_R$ point in 4-dimensions, 
one expects that there should be 11-dimensional uplifts of
4-dimensional flow equations satisfied in the $G_2$ point
corresponding to (\ref{four}). 

\vspace{.7cm}



\appendix

\renewcommand{\thesection}{\large \bf \mbox{Appendix~}\Alph{section}}
\renewcommand{\theequation}{\Alph{section}\mbox{.}\arabic{equation}}

\section{The 4-form 
field strength in frame basis }

One can read off the 4-forms from (\ref{c3}), 
(\ref{a3flow}) and (\ref{geometric})
and they are given in the frame basis as follows:
\bea
F_{1234}  & = &
\frac{3^{\frac{1}{4}}\left[c_{\mu}^2(-5+\cosh2\chi)+2\rho^8(-2+
c_{2\mu}+s_{\mu}^2 \, \rho^8\, \sinh^2\chi )\right]}
{\hat{L} \,  \rho^{\frac{4}{3}}\, \cosh^{\frac{2}{3}}\chi \, 
(c_{\mu}^2+\rho^8 \, s_{\mu}^2)^{\frac{4}{3}}}, \nonu \\
F_{579\,10} + i \, F_{567\,10}  & = &   \frac{3^{\frac{1}{4}} 
\, \rho^{\frac{8}{3}}\,(-1+ \rho^8) \,\sinh\chi}
{\hat{L} \, \mbox{sech}^{\frac{1}{3}}\chi   
\, (c_{\mu}^2+\rho^8 \, s_{\mu}^2)^{\frac{4}{3}}}\,s_{2\mu} \, 
 e^{i \delta} =-F_{568\,10} + i
\, F_{589\,10},
\nonu \\
F_{579\,11} + i \, F_{567\,11} & = &   - \frac{3^{\frac{1}{4}} 
\,  (3 + \rho^8)\, \sinh\chi}
{\hat{L} \,  \rho^{\frac{4}{3}}  \, \cosh^{\frac{2}{3}}\chi \, 
(c_{\mu}^2+\rho^8 \, s_{\mu}^2)^{\frac{1}{3}}}
\, e^{i \delta} =-F_{568\,11} + i \, F_{589\,11}
\nonu \\
& = &
-F_{67\,10\,11} + i \, F_{79\,10\,11} = -F_{89\,10\,11} - i \, F_{68\,10\,11},
\nonu \\
F_{1235} & = & \frac{3^{\frac{1}{4}} 
\, \rho^{\frac{8}{3}}\,\left[1+ \cosh2\chi + \rho^8\, (-3+\cosh 2\chi)
\right]}
{ \hat{L} \, \cosh^{\frac{5}{3}}\chi    
\, (c_{\mu}^2+\rho^8 \, s_{\mu}^2)^{\frac{4}{3}}} \, 
s_{\mu} \, c_{\mu}, \nonu \\
F_{4567} + i F_{4568} & = & -
 \frac{3^{\frac{1}{4}} 
\, (-3+\rho^8) \, \sinh 2\chi }
{ 2 \hat{L} \, \rho^{\frac{4}{3}}\, \cosh^{\frac{5}{3}}\chi      
\, (c_{\mu}^2+\rho^8 \, s_{\mu}^2)^{\frac{1}{3}}} \, e^{i\delta} 
=F_{4589} - i F_{4579} \nonu \\
&= &
-F_{468\,10} + i F_{467\,10}=F_{479\,10}+i F_{489\,10}, \nonu \\
F_{479\,11}+i F_{467\,11} & = 
& -\frac{3^{\frac{1}{4}} \,  \rho^{\frac{8}{3}} \, 
 \mbox{sech}^{\frac{5}{3}}\chi \, 
\sinh\chi \,\left[1+ \cosh2\chi +\rho^8 \,
(-3+\cosh2\chi) \right]}
{2 \hat{L}    
\, (c_{\mu}^2+\rho^8 \, s_{\mu}^2)^{\frac{4}{3}}} \, s_{2\mu} \,
e^{i\delta} \nonu \\
& = & -F_{468\,11} + i F_{489\,11}.
\label{F41}
\eea
For simplicity, we ignored the $r$-dependence on $\rho$ and $\chi$ in
the right hand side of (\ref{F41}). At the IR critical point, the only
first half of these survives, which is consistent with (\ref{F4}).

\section{The 4-form 
field strength in coordinate basis }

For convenience, let us present the 4-forms in coordinate basis.
First of all, the 3-form gauge field with 3-dimensional M2-brane
indices appearing in (\ref{a3flow}) provides the following two
4-forms. One  is given by
\bea
F_{1234} =\frac{3^{\frac{1}{4}}}{2 \hat{L} \, \rho^4}
e^{3A} \, \left[ 2\, c_{\mu}^2 \, \cosh^2\chi \, (-5 + \cosh2\chi) + 4
  \,  \rho^8 \, (-2+c_{2\mu}) \, \cosh^2\chi  +  \rho^{16} \, s_{\mu}^2 \, \sinh^2 2\chi
\right],
\nonu
\eea
where $r \equiv x^4$
and the other is 
\bea
F_{1235} = \frac{3}{2\rho^2} \, e^{3A} \, s_{2\mu} \, \left[ 1+ \cosh 2\chi
  + \rho^8 \, (-3+\cosh 2\chi)  \right],
\nonu
\eea
which vanishes at IR point and  $\mu \equiv x^5$.
The six $(45mn)$-components, in unusual notation, are given by
\bea
\frac{\sqrt{F G} \left( F_{4567}, F_{4568}, F_{4569}, F_{4578}, F_{4579},
    F_{4589} \right)}{
9\sqrt{3} \, \hat{L}^2 \, c_{\mu}^2\, \tanh \chi \, \rho^{-2} \, (-3 + \rho^8)}
= \left[ -\frac{1}{2} (\eta-\xi) c_{\delta}, -F
    s_{\delta},  -\eta  F  s_{\delta},
    G  s_{\delta},  
\xi  G  s_{\delta},  2 F  G  c_{\delta} \right], 
\nonu
\eea
which vanish at the IR-critical point,
and the sixteen $(4mnp)$-components are given by
\bea
&& \frac{\sqrt{F G} \left( F_{4678}, F_{4679}, F_{467\,10}, F_{4689},
  F_{468\,10}, F_{469\,10},
F_{4789}, F_{478\,10}, F_{479\,10}, F_{489\,10} \right)}
{ 9 \sqrt{3} \, \hat{L}^2 \, c_{\mu}^3 \, s_{\mu}\, \tanh \chi \, X^{-2} \,
\rho^6 \left[c_{\mu}^2(-2+\cosh 2\chi) + (-2+\cosh 2\chi -2\cosh^2 \chi \, 
s_{\mu}^2)\rho^8 +s_{\mu}^2\, \rho^{16} \right] }
\nonu \\
&& = \left[ (\eta^2-\xi^2)  s_{\delta},
 \eta \xi (\eta-\xi)  s_{\delta},
- \frac{1}{2}(\eta-\xi)  s_{\delta},
 2 \eta^2  F  c_{\delta}, 
 F \, c_{\delta}, 
\eta  F c_{\delta},
 -2 \xi^2  G  c_{\delta},
 -G  c_{\delta},
 -\xi  G  c_{\delta},
 2 F  G  s_{\delta}
\right]
\nonu
\eea
and
\bea
&& \frac{\sqrt{F G} \left( F_{467\, 11}, F_{468\, 11}, F_{469\,11}, F_{478\,11},
  F_{479\,11}, F_{489\,11} \right)}
{ 9 \sqrt{3} \, \hat{L}^2 \, c_{\mu}^3 \, s_{\mu}\, \tanh \chi \, X^{-2} \,
\rho^{-2} \left[3 c_{\mu}^2 + (1-2c_{2\mu}  +2\cosh^2 \chi \, 
s_{\mu}^2)\rho^8 +(-4 + \cosh2\chi )\, s_{\mu}^2 \, \rho^{16}\right] }
\nonu \\
&& = \left[ 
-\frac{1}{2}(\eta-\xi)\, s_{\delta},
F \, c_{\delta},
\eta \, F \, c_{\delta},
-G \, c_{\delta},
-\xi\, G \, c_{\delta},
2\, F \, G \, s_{\delta}
\right],
\nonu
\eea
where the functions $F, G$ and $X$ are given by (\ref{FG}) and 
(\ref{X}).
These  $(4mnp)$-components vanish at the IR-critical point also.
The sixteen $(5mnp)$-components are 
\bea
&& \frac{\sqrt{F G} \left( F_{5678}, F_{5679}, F_{567\,10}, F_{5689},
  F_{568\,10}, F_{569\,10},
F_{5789}, F_{578\,10}, F_{579\,10}, F_{589\,10} \right)}
{ 27 \cdot 3^{\frac{1}{4}} \, \hat{L}^3 \, c_{\mu}^4 \, \tanh \chi \, X^{-2} \,
\left[3c_{\mu}^2 + (3-2c_{2\mu})\, \rho^8-s_{\mu}^2\, \rho^{16}\right]}
\nonu \\
&&
= \left[ (\eta^2-\xi^2) s_{\delta}, 
 \eta \xi (\eta-\xi) s_{\delta},
 - \frac{1}{2} (\eta-\xi)\, s_{\delta},
 2 \eta^2  F  c_{\delta}, 
 F  c_{\delta},
 \eta  F  c_{\delta},
 -2 \xi^2  G  c_{\delta},
 -G  c_{\delta},
 -\xi  G  c_{\delta},
 2 F  G  s_{\delta}  \right],
\nonu
\eea
and
\bea
\frac{4\sqrt{F G} \left( F_{56711}, F_{56811}, F_{56911}, F_{57811},
  F_{57911}, F_{58911} \right)}
{27 \cdot 3^{\frac{1}{4}} \, \hat{L}^3 \, s_{2\mu}^2 \, \tanh \chi \, X^{-2} \,
(c_{\mu}^2 + 3\rho^8 +s_{\mu}^2\, \rho^{16})}
= \left[ -\frac{1}{2}(\eta-\xi) s_{\delta},
 F  c_{\delta},
 \eta  F  c_{\delta}, 
 -G  c_{\delta},
 -\xi  G  c_{\delta},
 2 F  G  s_{\delta}
\right].
\nonu
\eea
Finally,  the ten $(mnpq)$-components are given by
\bea
&& 
\frac{\sqrt{F G} \left( F_{678\,11}, F_{679\,11}, F_{67\,10\,11}, F_{689\,11},
  F_{68\,10\,11}, F_{69\,10\,11},
F_{789\,11}, F_{78\,10\,11}, F_{79\,10\,11}, F_{89\,10\,11} \right)}
{27 \cdot 3^{\frac{1}{4}} \, \hat{L}^3 \, c_{\mu}^3\, s_{\mu}\, \tanh \chi \, X^{-1} \,
(3 + \rho^8)} =
\nonu \\
&& 
\left[ -(\eta^2-\xi^2)  c_{\delta},
- \eta \xi (\eta-\xi)  c_{\delta},
\frac{1}{2} (\eta-\xi)  c_{\delta},
2 \eta^2  F s_{\delta}, 
F  s_{\delta},
\eta F  s_{\delta},
-2 \xi^2  G  s_{\delta},
-G  s_{\delta},
-\xi  G  s_{\delta},
-2  F  G c_{\delta} \right].
\nonu
\eea
Compared with the ones in Appendix A, the extra $F_{4569},
F_{4578},F_{469m},F_{478m}, F_{569m}, F_{578m}, F_{69m\,11}$ and $F_{78m\,11}$
components in coordinate basis are new.
The $F_{4589}, F_{489\,10}, F_{489\,11}, F_{589\,10},
F_{589\,11}$ and $F_{89\,10\,11}$ components are nonzero, 
but these are vanishing in the case of \cite{Ahn0909}.
For 4-forms with upper indices, there are nonzero components 
$F^{45m\,10}, 
F^{4m\,10\,11}$,  
and $F^{5m\,10\,11}$($m =6,7,8,9$)
and nonzero components with the same indices of lower 4-forms obtained
previously
except the following components 
$F^{4mnp}, 
F^{5mnp}$, and $
F^{mnp\,11}$($m,n,p=6,7,8,9$)
that are vanishing.


\end{document}